# Aleatory Architectures

Sean Keller, College of Architecture, Illinois Institute of Technology, Chicago, Illinois 60616

Heinrich Jaeger, James Franck Institute and Department of Physics,
The University of Chicago, Chicago, Illinois 60637

Abstract

Aleatory architectures explore new approaches and concepts at the intersection of granular materials research and architecture/structural engineering. It explicitly includes stochastic (re-) configuration of individual structural elements and suggests that building materials and components can have their own agency — that they can be designed to adapt and to find their own responses to structural or spatial contexts. In this Guest Editorial we introduce some of the key ideas and ask: Can there be design by disorder? What are the possibilities of material agency? Can we develop a vocabulary of concepts to interpret various orderings of chance? Several papers in this special issue then investigate these questions in more detail from a range of different scientific and architectural perspectives.

The centerpiece of the 2008 Olympics in Beijing was an enormous monument to a metaphor — a metaphor about order, hierarchy, design, and nature (and politics and power). The National Stadium (**Fig. 1**), ubiquitously referred to as the "Bird's Nest," was conceived by the Swiss architects Herzog & de Meuron as an unordered, non-hierarchical structure in which each element played an equally supporting role. As in an actual bird's nest, the clear overall form was to be arrived at through the accretion of (ostensibly) randomly-placed pieces. The compelling, though debatable, aesthetic and political performance of this structure has been described by one of us elsewhere [1]. What concerns us here is that the Beijing stadium remained *metaphorical*: not only were the massive pieces of steel laboriously bound together at each intersection (unlike their avian counterparts) but the structure itself was in fact much more mundane than it appears. In order to simplify the engineering and construction Herzog & de Meuron's initial idea of a non-hierarchical structure was replaced by a system of regularly spaced primary arches concealed within a "nest" of connecting pieces. Designed by one of the world's most important architectural practices for the world's most visible stage this "Bird's Nest" embodied in 42,000 tons of steel the image of a radically different architecture that it could not itself realize. Our question here is whether — by working out from the scales of physics and material science — this radically different architecture could actually be created.

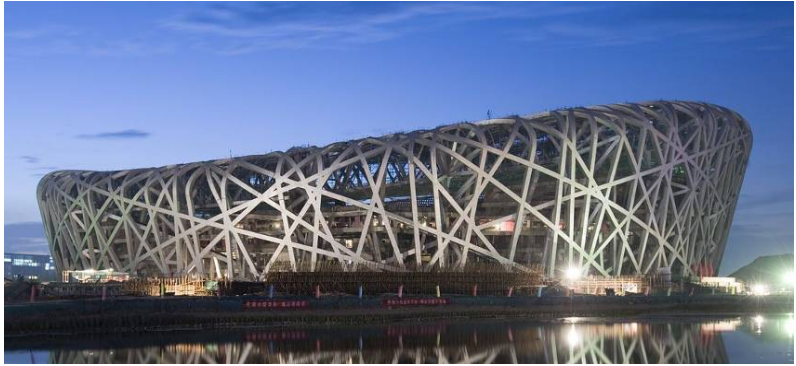

**Figure 1.** National Stadium in Beijing. Herzog & de Meuron, Chinese Architectural Design & Research Group, Arup, Ai Weiwei; 2002–8.

Architects have rarely allowed the undetermined to enter their design processes. Surely this is a result of the close — almost axiomatic — identification of design with order(-ing): to design is to give order, and this seems to mean to eliminate chance. Contrast this with applications using granular matter, which rely on statistical descriptions of its properties. As a random aggregate, granular matter does not require the *a priori* placement of individual particles. Their arrangement is not regular and ordered, but irregular and disordered. Stresses propagate not through regular, fixed supports, such as columns, but along a vein-like network of paths that self-configures, and potentially re-configures, in response to external loads [2-6] (**Fig.2**). Thus, the normal requirements and methods of architects and building engineers appear antithetical to the options provided by granular matter.

Our modern built environments reflect this. Carefully planned and executed design and construction methods in concert with advanced building materials have enabled skyscrapers, vaults, cantilevers, and bridges of remarkable sophistication. On the other hand, most structures made of granular matter are comparatively simple. The predominant focus on ubiquitous types of granular media, such as sand or gravel, means that the range of resulting forms and properties is limited. Freestanding structures such as piles, mounds, or dams have not evolved much in form over the millennia because their general shape is tied to a relatively small available range in the angle of repose of typical granular particles. Similarly, the range of accessible properties, such as porosity, is small for standard granular matter.

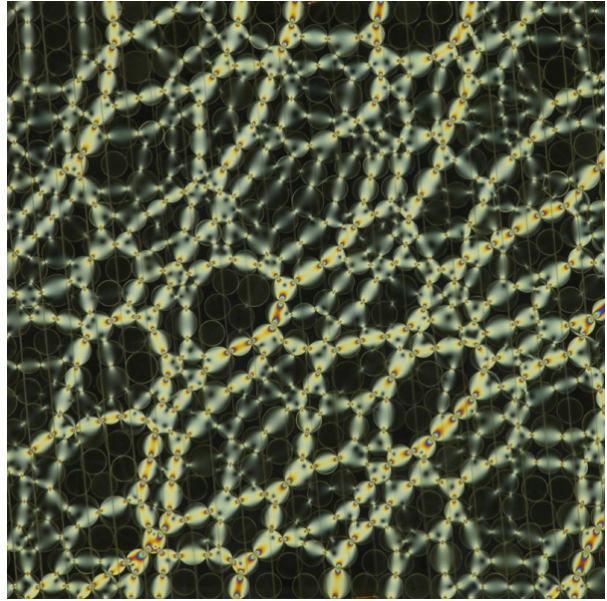

**Figure 2.** Fabric of force chains inside a granular material, visualized here by stress-induced birefringence in a two-dimensional packing of discs. A load is applied uniformly across the top, which is then sheared. Large local stresses give rise to bright spots in this image. Image credit: R. P. Behringer, Duke.

Yet, recent research has started to overcome some of these limitations and there are a number of properties of granular matter that offer opportunities not available with other materials. Findings from a surge of recent studies using more complex, non-spherical particle shapes have made it possible to start designing granular aggregates with target properties previously out of reach — such as aggregates that are not only stronger or tougher, but that combine high porosity with high strength, or that are self-strengthening under load. In addition, much has been learned about the unique ability of granular matter to transform, reversibly, between free flowing "unjammed" and rigid "jammed" configurations. Control of the jamming transition enables tuning of aggregate stiffness and strength over a wider range than is accessible with most other materials. The fact that granular matter is disordered on either side of the jamming transition also implies an ability to self-heal that does not exist for structures whose strength depends on precise placement of the constituent elements.

Independently, architects have begun to explore previously uncharted territory, asking questions such as: What if design or construction methods could include rather than exclude an element of chance? Can randomness and structural disorder, in fact, lead to forms that perform better or that are richer in effect and meaning?

Aleatory architecture explores concepts at the intersection of granular materials research and architecture/structural engineering. We coined the term to imply a new approach that explicitly includes stochastic (re-) configuration of individual structural elements — that is to say "chance." Aleatory architecture disrupts the traditional assumptions about the authority of the architect as planner as well as the typical hierarchy of the design process. Aleatory architecture suggests that building materials and components can have their own agency — that they can be "designed" to adapt and to find their own responses to structural or spatial contexts. In this way the very meaning of "design" is brought forward for reconsideration (though not abandonment): Can there be design by disorder? What are the possibilities of material agency? Can we develop a vocabulary of concepts to interpret various orderings of chance?

In classical Latin, *ālea* refers to a die or dice, and so *āleātōrius* means "connected to gambling." In modern usage, "aleatory" became an important concept in mid-twentieth century avant-garde art—especially music, where many varieties of indeterminate composition and performance were explored. For example, John Cage's seminal *Music of Changes* (1951) demonstrates how chance devices of some sort—rolling dice, flipping coins, shuffling cards, dropping sticks—can be used to generate a musical score, thereby weakening the role of the composer (as well as the preconceptions of the performer and listeners). What we consider here is how a similar exploration of chance's role within composition might take place in architecture—emphasizing that this cannot be a simple copying of aleatory techniques already used in other fields, but requires a careful translation into the specific possibilities and restrictions of contemporary building.

A confluence of factors has made it possible to begin thinking about aleatory methods in architecture. On the technical side are advances in computation, material science, and robotics. These fields have begun to embrace the tolerance, robustness, and adaptivity that disorder can confer. There is an emerging recognition that one of the most refined design processes we know of —evolution— exploits variability and heterogeneity for exactly those benefits. Aleatory design also exemplifies a mature, "post-hype" use of computational tools: moving away from exclusively digital simulations to savvy hybrids that make the best use of both computational and physical modeling. Culturally, this approach can be seen as one result of the long dissolution of architectural modernism and its underlying (and ultimately classicist) search for aesthetic stability. Aleatory architecture challenges architecture's prejudices for totalizing order and control.

In keeping with this open, hybrid attitude, aleatory design is not intended to replace current building methods. Instead, it is meant to explore new options, expanding the range of possibilities. In many, perhaps most, cases aleatory construction may be combined with more traditional building elements such as spanning plates, tensile components, and enclosing membranes. Each approach can be used where it is most effective. What aleatory architecture hopes to do is greatly expand a range of approaches that, up to this point, have been too narrowly restricted.

For example, what if speed of construction and low material weight become more important than structural perfection or permanence? Speed certainly is a prime benefit of any approach based on granular matter: since no regular, ordered arrangement is required, granular structures can be simply poured into place. Using suitable particle shapes, low-packing-density and thus potentially lightweight structures can be envisioned. Since the constituent elements connect solely by interlocking and/or friction, such structures are not only formed very quickly, but also reconfigured and taken apart easily. Thus an aleatory approach based on granular matter challenges the traditional architectural notion of planning for structural permanence.

The umbrella term "granular matter" covers an enormous range of materials with different particle types and sizes, from fine powders to huge boulders [7,8]. Similarly, jamming transitions exist for all kinds of materials and particles [5,9-14]. (In fact, granular jamming-based concepts are starting to be applied to systems comprised of ever smaller constituent elements, including nanoparticles, for example for encapsulation of liquid drops [15] or for creating high-efficiency battery electrodes [16].) With aleatory architecture we focus on the scale of large manmade structures, whose aesthetic as well as physical properties emerge from the requirement that they can interface with the human figure and with the rest of the human-built environment.

In principle this includes landscape and infrastructure elements such as harbor breakwaters or gravel beds for railroads. All of these structures rely on the many unique advantages of granular matter, including its porous nature that enables quick drainage, its adaptability to load variation, the option to easily reconfigure and reshape the overall form, and of course the fact that simple piling, pouring or bulldozing can deploy the material quickly.

Yet with aleatory architecture we intend to take a significant further step. What we are exploring here is whether new ideas could lead to new architectural structures based on granular, jamming-based construction. Thus, on the level of science and engineering we ask: How far can we push an aleatory conception of construction? What are the technological challenges and the eventual limits? And on the architectural level we ask: What is the emerging aleatory aesthetic and how can it be turned to architectural purposes?

Our aim with this article is to outline the context for these questions by connecting advances in granular materials physics and engineering with recent pertinent approaches on the architectural side. In this issue of *Granular Matter* four papers by architects and scientists then provide an overview of current research activities at the intersection of these fields.

**Challenges and Opportunities**

From the perspective of science and engineering, there are several aspects that make this an exciting area of exploration.

A longstanding, fundamental challenge in material science has been predicting the macroscopic aggregate properties of a local microstructure that is so highly disordered and irregular that it can only be described in statistical terms. This applies at all scales, whether the constituent elements are of molecular size in an amorphous glassy material or are randomly piled twenty-ton boulders in a harbor breakwater. For the situation of interest here, in which individual particles are large and thermally driven particle motion becomes irrelevant, the concept of jamming provides an excellent framework for advancement [10-12,14]. Jamming describes the behavior of a granular aggregate using only two key parameters: the degree of particle "crowding," as measured by the average packing fraction, and the amount of applied stress. A granular system is jammed and mechanically stable when the particle packing fraction is sufficiently high and as long as any applied stress remains sufficiently low.

The challenge is that these conditions apply in a statistically averaged manner, and there can be significant fluctuations, inducing instability even if the parameters nominally fall within a stable regime. Therefore, one might worry about whether it is possible to meet the stringent stability requirements imposed by engineering standards on current architectural structures. But to us this misses the point: aleatory architecture based on granular jamming is not meant to compete in this existing arena. As with all new approaches it will need to find its niche and, reciprocally, engineering standards will need to adjust to these new techniques. The more relevant questions relate to whether jammed granular aggregates are amenable to generate predictive capabilities in the first place. For example: Can we predict how much, on average, a granular aggregate comprised of arbitrarily shaped particles will strain under a given applied load? Or, on a more detailed level: Can we predict the likelihood that an observed strain will deviate from that average by a certain amount? The answers to questions like these are beginning to become "yes."

Again advances in the last few years have been essential to greater understanding of these systems. In the past, the structural complexity of large amorphous aggregates has meant that large systems could typically be simulated only if they were comprised of simple, usually spherical, particles. Until recently, the difficulty in fabricating non-spherical particles has meant that physical experiments typically had to resort to naturally occurring shapes, as in sand or gravel, or particles like polyhedra of different shapes, discs, or rods that could easily be procured in large quantities. This has all changed recently. Modern computational tools for granular systems have now reached a level of power and sophistication that makes it possible to calculate the properties of large aggregates of arbitrarily shaped particles, and fabrication methods such as 3D printing have enabled a new generation of experiments that can explore the role of complex particle shape much more systematically.

However, to be useful for architectural design it is not sufficient to be able to predict the behavior resulting from a given collection of materials or particles. A successful design process, in architecture or elsewhere, has to be able to follow the inverse path: starting from overall, large-scale target properties that are to be reached and working out the required lower-level ingredients together with an assembly pathway.

Granular matter poses at least two intriguing challenges in this regard. One is the fact that we have to design an irregular, amorphous aggregate—i.e. the design process has to consider explicitly the statistical description of the local particle configurations. The second is the fact that the boundary conditions as well as the processing conditions are critically important for the behavior of a granular aggregate, much more so than is the case for ordinary materials such as glass or steel. By boundary conditions we mean, in particular, the degree of overall confinement due to external pressure on the aggregate; and by processing conditions we refer to the manner in which the material is poured, or whether or not it is agitated (for example by tamping or vibrations). It is therefore imperative to embed these aspects into the design process, in addition to any considerations relating to the particles' shape and material properties [14].

For granular matter, the standard approach to these "inverse problems" that are at the core of design is still based largely on trial and error experimentation. This has worked for simple architectural forms, such as in soil mechanics, where clear design rules have emerged from a large body of knowledge built up over decades. Aleatory architecture pushes into territory where little empirical knowledge exists and ready-to-use design rules will need to be developed. This absence of rules makes it particularly fertile ground for computer-aided design and optimization methods. In fact, we would argue that such methods will become indispensible.

In this context, there is one development that has made large inroads into manufacturing and is now also starting to be applied on architectural scales [17,18]. This is the advent of robotic on-demand fabrication techniques and 3D-printing. The combination of such techniques with advanced computation capabilities has made it possible to design and then fabricate highly customized building components. This approach opens up the accessible range of forms and shapes, allowing for almost arbitrary variability without increasing fabrication cost, while at the same time enhancing fabrication precision. It is important to note, however, that this approach only works with extensive pre-planning. While the form of the physical material is liberated from the straightjacket of (cost-driven) standardization, the burden of having to pre-determine the exact placement of every bit of material is still there; it simply has been shifted from hardware into the virtual realm of software and computation.

Jamming-based aleatory architecture introduces a radically different perspective. The fact that individual particles configure and re-configure in response to external forces like gravity means the granular material determines all local adjacencies and arrangements autonomously. This "material computation" [19] occurs in the actual physical structure, i.e. the hardware, and it occurs in real-time during fabrication or later in response to load changes.

An aleatory approach thus aims to achieve form without requiring pre-planned structural adjacencies. If properly chosen, the material will adapt its configuration automatically to become load-bearing (in contrast to traditional arches that require carefully placed keystones). In a sense, the most difficult aspect of granular matter, namely dealing with the local disorder in any *given* particle packing (as opposed to dealing with statistical averages), is taken care of by the material itself.

As a result, pre-planning is freed from considering the local structural detail and complexity that goes hand in hand with disorder. Instead, the main task now becomes generating the proper particle shapes as well as the overall boundary and processing conditions to guarantee that the desired target structure will be mechanically stable when realized. In this way aleatory architecture suggests the potential of focusing design efforts on a middle ground, for which particles are not simply pre-given and handled as masses, as in traditional aggregate materials, but are also not individually designed and positioned, as parts, as in traditional mechanical design.

From an architectural perspective, the aleatory aesthetic offers a number of opportunities. One is the high degree of geometric and textural detail it makes possible and its inherent heterogeneity. The seemingly random, stochastic particle configurations of jammed granular assemblies openly defy notions of pre-ordained regularity and in this way introduce quasi-organic qualities at new and larger scales than have previously been possible. Recent advanced architecture has an acknowledged tendency to create new relationships between structure and ornament, relationships in which the traditional hierarchy between the two (structure primary; ornament secondary) is collapsed, often by creating single systems that are both structural and ornamental [20,21]. Aleatory architecture suggests new ways of working in this direction as irregularly patterned assemblies of elements become structurally significant themselves.

A key point, however, is that these adjacencies and configurations are not arbitrary: they emerge as a consequence of physical requirements for stability, such as force balance, endowing them with structural purpose. The resulting patterns challenge our notions of geometrical order, but are nonetheless quite "orderly" outcomes of physical forces. Aleatory architecture thus opens up new ways of thinking about received concepts such as structural clarity and material expression, thereby challenging modernism's association of geometric simplicity with material honesty. Recall here Louis Kahn's famous admonition to "ask the brick what it wants to be." Who really believes that the brick "wants to be" arranged into the simple Euclidian shapes of squares, circles, and triangles that were Kahn's architectural vocabulary? Surely those forms were imposed on every brick that was set, by hand, according to a drawing from Kahn's office. In contrast, aleatory architecture posits greater agency within the construction process for materials elements and fundamental physical forces.

**A Bit of (Recent) History and Some Examples**

We are just beginning to realize this concept as a new architectural approach. The papers in this issue demonstrate first steps by focusing on basic architectural elements or protostructures, such as walls, columns, or domes. In order to go beyond sloped mounds of granular material and to construct tall freestanding structures by random assembly without bonding or fastening, some form of confinement is needed to keep the particles in place. In general, there are two options: either an additional structural component is needed, such as a membrane or other support over the outer surface of the granular aggregate, or the particles self-confine, for example via entanglement.

Enclosing granular material in a non-porous membrane provides a means to apply a large uniform confining pressure by evacuating the interior, thereby driving the material into a deeply jammed state. Over a wide range by the confining pressure the stiffness and strength of the aggregate can be controlled, from highly malleable to fully rigid [22]. This makes it possible to bend or reconfigure the material before rigidly jamming it to lock in the final shape, an approach that has also been used for soft-robotics applications [23-27]. There have been a number of exploratory architectural projects based on such vacuum jamming (see **Fig. 3**), including small domed shelters and carports as well as proposals for reconfigurable housing by the architects John Gilbert and William Hanna (Queens University, Belfast) in 1970 [28], and more recently enclosures and sustainable material structures at the Eindhoven University of Technology [29,30] as well as light, quickly deployable bridges tested by Ulrich Knaack's team at the Technical University Delft [31].

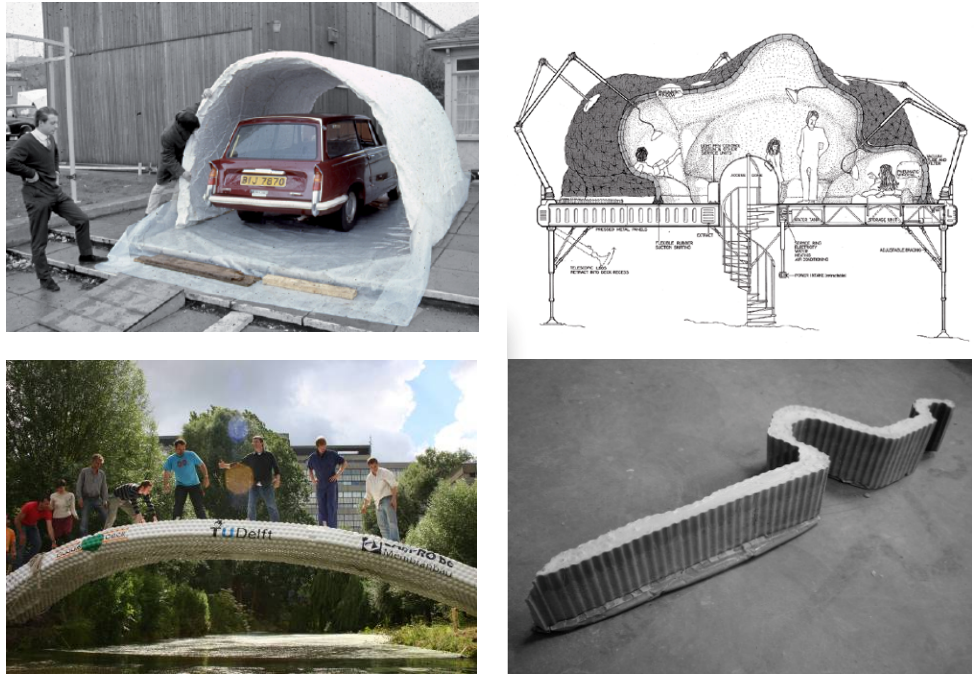

**Figure 3.** Examples of structures formed using granular materials together with vacuum jamming. (a, b) car port and reconfigurable 'living room' by John Gilbert and William Hanna. (c) Bridge from the "Deflateables" project. (d) Cast concrete structure, shaped by using vacuumatics formwork.

In the architectural context, there is also an important complementary application for such "vacuumatics": the use of jammed structures as reconfigurable molds for casting the complexly curved free-form shapes that have been part of the architectural vocabulary since the advent of high-strength concrete. What is typically forgotten after a concrete roof or bridge has been erected, is the considerable time and effort that went into producing the formwork that supported the concrete until it cured and became load bearing — formwork that is often discarded after a single use. Vacuum jamming of granular media provides an exciting alternative because it can be easily shaped as well as reused, with the added benefit that embedded particles (or other solid objects) protruding through the membrane allow for texturing of the concrete surface [30,32-34]. The article by Frank Huijben on page [**to be inserted**] in this issue explores this further, focusing on flexibly reconfigurable vacuumatic formwork for the casting of topology-optimized concrete structures.

Aggregates jammed through vacuumatics can be as thin as a few particle diameters and thus almost sheet-like (**Fig. 3a-c**). However, in many instances the requirement of a vacuum-tight membrane is a limitation. Unless transparent, such membrane also obscures the material inside. More immediate access to the raw materiality of a granular aggregate is provided by alternative types of confinement.

One of these is an open mesh. The gabion — a wire basket filled with stones — has long been used for creating steep slopes for retaining walls and embankments. Similarly, by providing confinement and reinforcement, geogrids have been used to stabilize slopes steeper than would be possible with the natural angle of repose of gravel or soil [35].

A more recent extension of gabion-type wire mesh confinement are stone-filled panels that perform as vertical wall elements. Applications range from fences to whole buildings. A significant example is the 50,000 sqf structure for the Dominus winery in Napa Valley, CA, designed by the firm of Herzog & de Meuron (Basel) and completed in 1998 (**Fig. 4**).

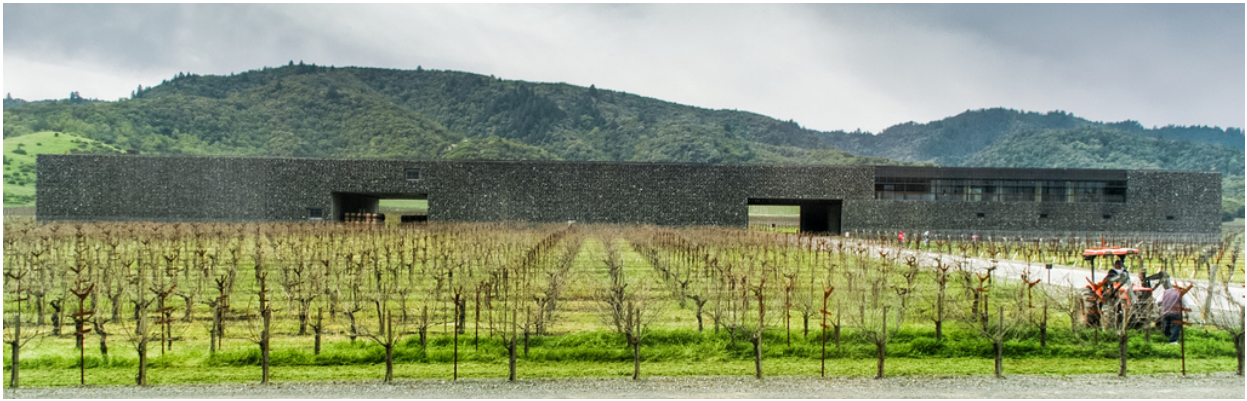

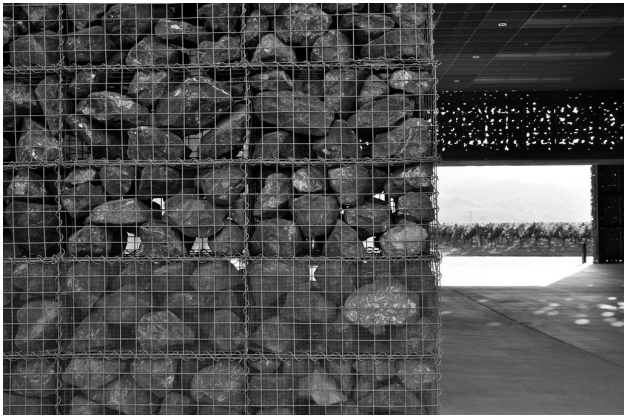

**Figure 4.** Dominus Winery, Yountville, CA. Herzog & de Meuron, 1995-98.

Herzog & de Meuron's gabion-based approach succeeded in meeting several key design objectives. The natural stone material provided the desired integration with the surrounding vineyard and landscape. The disordered packing of the stones within the gabions was made sufficiently porous to let sunlight filter through, producing a very important aesthetic effect. At the same time, the large thermal mass of the stones provided efficient temperature regulation of the structure's interior, cooling it during the day and keeping it warm at night. Interesting tensions emerge here from the use of the most traditional of materials (irregular natural stones) and methods (pouring or random packing) to create a unique contemporary building that outperforms standard means of construction. The result in this instance is a structure that is technologically innovative, ecologically sensible, and aesthetically rewarding.

Nonetheless, confinement via the wire basket is still quite visible. It is also structurally significant and therefore the shape of the basket determines the overall form of the wall segment. Changes in this form would require reshaping of the basket.

A completely novel concept for supplying confinement has recently been introduced by the group of Gramazio Kohler Research at ETH Zurich in collaboration with the Self-Assembly Lab of MIT. Their approach too uses irregularly shaped small rocks or gravel, i.e, it focuses on readily available types of granular matter, with the added benefit that the particles could be made from recycled material such as glass. However, instead of employing pre-fabricated gabions, they achieve confinement by adding thin, flexible string *in situ* while the aggregate is being built up (**Fig. 5**). Together with strongly frictional particle-particle interactions (the particle material is chosen for high friction) the string provides the tensile forces required to hold the aggregate together and enables vertical walls as well as slender columns (see the paper on page [**to be inserted**] in this issue).

While there is no pre-planning as far as the granular material is concerned, the string placement is not random, but by careful design. Importantly, the final aggregate is mechanically stable only in the regions patterned with string; particles in regions that remain unpatterned will simply fall off under gravity. This approach - pursued in collaboration with the research group of Hans J. Herrmann of ETH Zurich (see the paper on page [**to be inserted**] in this issue) opens up fascinating new options: not external confinement but a string running through the interior of a large granular aggregate defines the final form. The process resembles 3D-printing by local laser-fusing of metal powder beds [36], with the string taking on the task of connecting particles, i.e. providing just enough tensile strength that the aggregate holds together. In its scaled up version, which uses large industrial robots to pour the gravel and lay the string pattern, Gramazio and Kohler thus call their technique "3D rock printing."

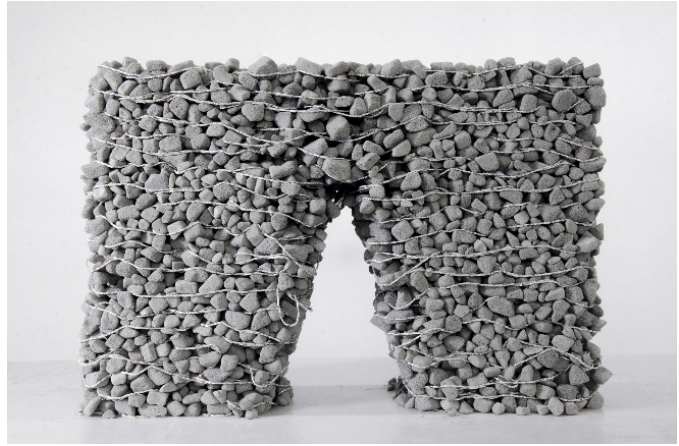

**Figure 5.** Prototype wall element with arch. Gramazio & Kohler Research, 2014.

Once the confinement by external means, or added features such as string, is removed, the results tend to be pile-shaped forms, at least for standard granular media like sand or gravel. This could be viewed as the extreme limit of aleatory architecture: just the material in all its rawness plus the laws of physics. No material refinement on the one hand, and no preplanning or architectural intervention on the other, with the aggregate computing its configuration autonomously. This absence of preplanned form is the radical platform advanced by Garret Ricciardi and Julian Rose, two young architects who founded the New York firm Formlessfinder (**Fig. 6**). For them the "formless," a philosophical concept introduced by Georges Bataille, emphasizes the inherently physical nature of a material, as opposed to the particular shape or form it might attain through architectural intervention: "Form suppresses material and tends to either idealize architectural materials or dematerialize architecture altogether" [37].

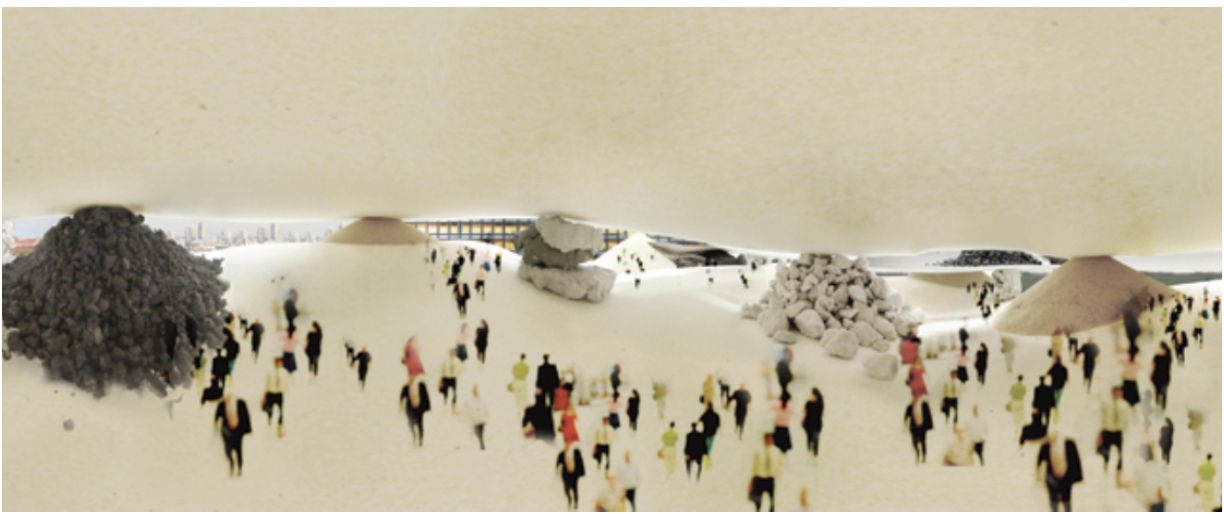

**Figure 6.** Concept study *Load Test*. Formlessfinder, NY, 2010. From Ref. 37.

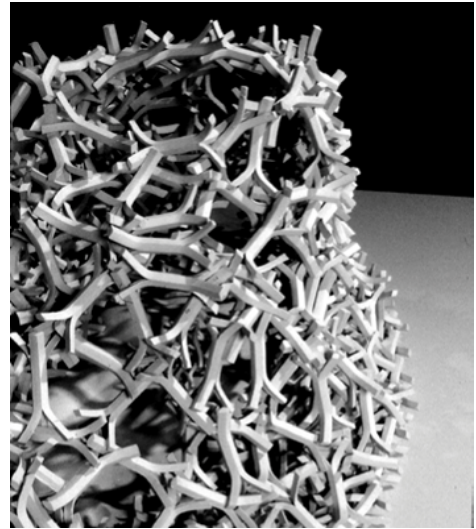

**Figure 7.** Scaffold constructed with *Tumbling Units*. Kentaro Tsubaki, 1997. From Ref. 38.

If we embrace the aleatory aspects and the materiality of the formless, yet at the same time want to explore a wider range of structural possibilities, one option is to move to particle shapes that produce self-confinement. These are shapes specifically picked to enable behavior not possible with typical naturally occurring or industrially produced bulk granular materials such as sand or gravel. They achieve self-confinement through interlocking or entanglement.

One of the earliest attempts to investigate the resulting visually complex configurations for their architectural potential was the "tumbling units" project by Kentaro Tsubaki, then at Tulane [38]. Tsubaki's tumbling units were particles he designed so they would interlock when poured and produce low-density, highly porous scaffolds. However, to a large part his interest was in exploring the potential for structural stability via geometry and friction, and less in pushing the aleatory aspects of random aggregates. Thus, models as shown in **Fig. 7** are still constructed rather than poured.

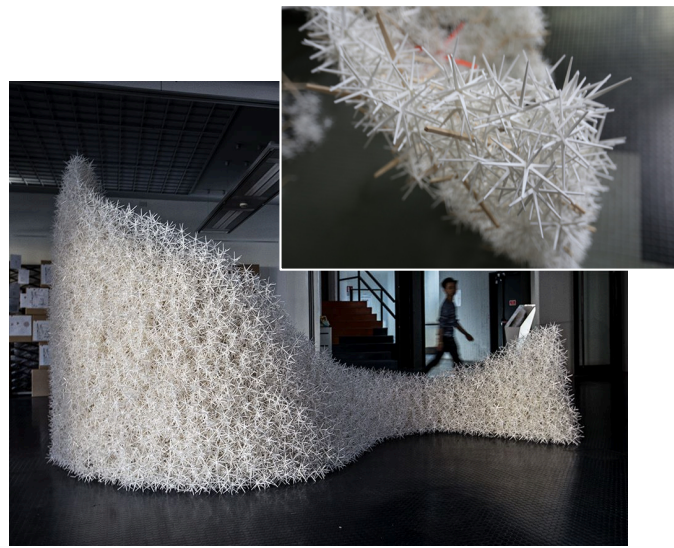

The first architectural structures that fully embraced aleatory aspects of construction appeared in a series of projects by Achim Menges and Karola Dierichs (University of Stuttgart), who developed this into an approach they termed "aggregate architecture" [19,39] (**Fig. 8**). These projects explored X-shaped particles (with the lower half rotated out of plane), various three-dimensional stars, and recently also hooked shapes (**Fig. 8**), see their article on page [**to be inserted**] in this issue. Their work beautifully demonstrates that aleatory methods, with suitably chosen, self-confining particles, can produce essentially all of the basic architectural protostructures, including walls, arches and domes. From the basic science perspective, the highly non-convex particle shapes pose important questions about stress transmission through the aggregate, an aspect recently investigated in collaboration with Bob Behringer's group at Duke University (see the article on page [**to be inserted**]).

**Figure 8.** *Aggregate Architecture*. Inset shows detail. Karola Dierichs and Achim Menges, ICD Stuttgart I ActLab Milan - Milan Architecture Weeks 2015

One interesting aspect running through this research has been the introduction of robotic methods to aleatory construction. The powerful possibilities emerging from the use of industrial robotic manipulators are at the core of the work of Gramazio Kohler Research [17] — as in the deployment of the string patterns — while Dierichs and Menges use robots to control the pouring of particles. Robotic methods such as additive manufacturing will also become increasingly important for the creation of more complex particle types. Looking ahead, there clearly will be a tension between, on the one hand, the apparent simplicity of the aleatory construction process coupled with the ubiquity and low cost of standard granular media, and, on the other hand, new structural and aesthetic possibilities achieved with high-end robotic technology and more complex and thus potentially more costly particle shapes.

In this context, an important option may be to search for the simplest particle shape that satisfies a given design target, and a basic architectural protostructure such as a slender column might make an excellent initial target. For example, we might ask: Is it possible to achieve a column with an aspect ratio (height/diameter) exceeding 10? For such a column to be stable, stresses have to be transmitted predominantly vertically downward, and any remaining horizontal components must be balanced by friction and/or tensile forces, if available. Particle types that easily produce shear bands clearly will not work. Ideally, one might want particle types that produce aggregates with a very small, nearly zero Poisson ratio. In the absence of additional means for confinement, this points toward self-confining shapes that strongly interlock or entangle.

One particle type that fulfills this condition is a flexible chain. Aggregates of chains composed of flexibly joined spheres have recently been shown to exhibit very strong strain stiffening. In contrast to ordinary granular media, which weaken under compression, these aggregates become significantly stiffer once the chains are long enough to form loops and entangle [40]. It is intriguing that such aggregates of chains, or "granular polymers," are stronger than aggregates formed with almost any other particle type, yet exhibit a much higher porosity and thus lower weight [40-42]. More complex chains, composed of non-spherical objects and with different types of connectors between adjacent particles, are investigated in the article by Tibbits et al. on page [**to be inserted**] of this issue.

Another option are rigid particles with hook-like shapes that enable interlocking. Work by Scott Franklin (Rochester Inst. of Technology), Dan Goldman (Georgia Tech), and coworkers has investigated U-shaped, "staple"-type particles [43,44]. This research path was originally motivated by the extended, yet remarkably strong, structures formed by fire ants, in which groups of 100,000 and more insects are held together by interlocking limbs and mandibles. However, aggregates of these particles are typically so well connected that they are hard to disassemble once formed. This may be advantageous for (semi-) permanent structures, but it negates one of the more intriguing aspects of aleatory architecture: easy recyclability and reusability. For example, just as a knitted sweater unravels easily, pulling the string in one of the structures formed by Gramazio Kohler Research's "3D rock printing" method causes a stable loadbearing form to turn into a pile of particles ready for immediate reuse.

In the lab of one of us, we investigated a particle variant in which one of the "arms" is rotated 180 degrees. This Z-shaped "crankshaft" particle retains much of the strong interlocking and strain-stiffening under load of the U-shaped variety, but has the benefit of easier disassembly when the load is removed. This makes it possible to envision structures that are incompatible with conventional wisdom: they are stable under load but disintegrate when the load is removed (see the article on page [**to be inserted**]).

Finally, in all of the work described the particles were assumed to retain a fixed shape throughout the construction and mechanical loading process. As an intriguing next step we can imagine particle types that actively respond to variations in the (local) environment and react by changing their shape or behavior. A first glimpse of the emerging possibilities for this strain of aleatory architecture is presented in the paper by Dierichs and Menges on page [**to be inserted**].

**Acknowledgements**

We would like to thank all authors who contributed to this special issue of *Granular Matter* for many illuminating discussions. HMJ acknowledges support from the National Science Foundation through grant CBET-1334426 and from the University of Chicago through an Arete Vision grant.

**References**

1 Keller, S.: Bidden City (Beijing Olympics). Artforum Int **46**, 137-145 (2008).
2 Radjai, F., Wolf, D. E., Jean, M. & Moreau, J. J.: Bimodal character of stress transmission in granular packings. Phys. Rev. Lett. **80**, 61-64 (1998).
3 Liu, C. H. *et al.*: Force Fluctuations In Bead Packs. Science **269**, 513-515 (1995).
4 Sanfratello, L., Fukushima, E. & Behringer, R. P.: Using MR elastography to image the 3D force chain structure of a quasi-static granular assembly. Gran. Matt. **11**, 1-6 (2009).
5 Bi, D., Zhang, J., Chakraborty, B. & Behringer, R. P.: Jamming by Shear. Nature **480**, 355-358 (2011).
6 Kondic, L. *et al.*: Topology of force networks in compressed granular media. Europhys. Lett. **97**, 54001 (2012).
7 Jaeger, H. M., Nagel, S. R. & Behringer, R. P.: Granular solids, liquids, and gases. Rev. Mod. Phys. **68**, 1259-1273 (1996).
8 Andreotti, B., Forterre, Y. & Pouliquen, O.: *Granular Media: Between Fluid and Solid*. (Cambridge University Press, 2013).

9  Cates, M. E., Wittmer, J. P., Bouchaud, J. P. & Claudin, P.: Jamming, force chains, and fragile matter. Phys. Rev. Lett. **81**, 1841-1844 (1998).
10 Liu, A. J. & Nagel, S. R.: *Jamming and rheology : constrained dynamics on microscopic and macroscopic scales*.  (Taylor & Francis, 2001).
11 Liu, A. J. & Nagel, S. R.: The Jamming Transition and the Marginally Jammed Solid. Annual Review of Condensed Matter Physics **1**, 347-369 (2010).
12 van Hecke, M.: Jamming of soft particles: geometry, mechanics, scaling and isostaticity. Journal of Physics-Condensed Matter **22**, 033101 (2010).
13 Reichhardt, C. & Reichhardt, C. J. O.: Aspects of jamming in two-dimensional athermal frictionless systems. Soft Matter **10**, 2932-2944 (2014).
14 Jaeger, H. M.: Toward Jamming by Design. Soft Matter **11**, 12-27 (2015).
15 Cui, M. M., Emrick, T. & Russell, T. P.: Stabilizing Liquid Drops in Nonequilibrium Shapes by the Interfacial Jamming of Nanoparticles. Science **342**, 460-463 (2013).
16 Srivastava, I., Smith, K. C. & Fisher, T. S.: Shear-induced failure in jammed nanoparticle assemblies. AIP Conference Proceedings **1542**, 86 (2013).
17 Gramazio, F., Kohler, M. & Willmann, J.: *The Robotic Touch - How Robots Change Architecture*.  (Park Books, 2014).
18 DUS_Architects: *3D Printed Canal House*, <http://3dprintcanalhouse.com> (2015).
19 Dierichs, K. & Menges, A.: Aggregate Structures: Material and Machine Computation of Designed Granular Substances. Architectural Design **82**, 74-81 (2012).
20 Moussavi, F. & Kubo, M.: *The Function of Ornament*. (Actar, Harvard Graduate School of Design, 2006).
21 Rappaport, N.: Deep Decoration. 30/60/90 Architectural Journal **10**, 95-105 (2006).
22 Athanassiadis, A. G. *et al.*: Particle shape effects on the stress response of granular packings. Soft Matter **10**, 48–59 (2014).
23 Brown, E. *et al.*: Universal robotic gripper based on the jamming of granular material. Proceedings of the National Academy of Sciences of the United States of America **107**, 18809-18814 (2010).
24 Steltz, E., Mozeika, A., Rembisz, J., Corson, N. & Jaeger, H. M.: Jamming as an Enabling Technology for Soft Robotics. Proc. SPIE 7642, Electroactive Polymer Actuators and Devices (EAPAD), 764225 (2010).
25 Follmer, S., Leithinger, D., Olwal, A., Cheng, N. & Ishii, H.: Jamming user interfaces: programmable particle stiffness and sensing for malleable and shape-changing devices, in Proc. *25th annual ACM symposium on User interface software and technology*. 519-528 (Assoc. for Computing Machinery, 2012).
26 Amend, J. R., Brown, E., Rodenberg, N., Jaeger, H. M. & Lipson, H.: A Positive Pressure Universal Gripper Based on the Jamming of Granular Material. IEEE Transactions on Robotics **28**, 341-350 (2012).
27 Stanley, A. A., Gwilliam, J. C. & Okamura, A. M.: Haptic Jamming: A Deformable Geometry, Variable Stiffness Tactile Display using Pneumatics and Particle Jamming, in Proc. *IEEE World Haptics Conference*. 25-30 (IEEE, 2013).
28 Gilbert, J., Patton, M., Mullen, C. & Black, S. Vacuumatics, 4th year research project. (Department of Architecture and Planning, Queen's University, Belfast, 1970).
29 Huijben, F. A. A.: *Vacuumatics: Vacuumatically Prestressed Reconfigurable Architectural Structures* Master's thesis, Department of the Built Environment, Eindhoven University of Technology, (2008).

30  Huijben, F. A. A.: *Vacuumatics: 3D Formwork Systems* PhD thesis, Eindhoven University of Technology, (2014).
31  Knaack, U., Klein, T. & Bilow, M.: *Deflateables / Imagine 02*. 128 (010 Publishers, 2008).
32  Huijben, F. & van Herwijnen, F.: Vacuumatics; shaping space by "freezing" the geometry of structures, in Proc. *Int'l Conference on Tectonics: Making Meaning 2007*. 1-11 (2007).
33  Huijben, F. & van Herwijnen, F.: Vacuumatics: vacuumatically prestressed (adaptable) structures, in Proc. *6th International Conference on Computation of Shell and Spatial Structures IASS-IACM 2008 "Spanning Nano to Mega",* eds. J. F. Abel & J. R. Cooke. 1-4 (Multi Science Publishing, Brentwood, UK, 2008).
34  Huijben, F., van Herwijnen, F. & Nijsse, R.: Vacuumatics 3D-Formwork Systems: Customised Free-Form Solidification, in Proc. *Structural Membranes 2009 (International Conference on Textile Composites and Inflatable Structures),* eds. E. Onate & B. Kröplin. 1-4 (2009).
35  Matsuo, O., Tsutsumi, T., Yokoyama, K. & Saito, Y.: Shaking table tests and analyses of geosynthetic-reinforced soil retaining walls. Geosynthetics International **5**, 97-126 (1998).
36  Gagg, G., Ghassemieh, E. & Wiria, F. E.: Effects of sintering temperature on morphology and mechanical characteristics of 3D printed porous titanium used as dental implant. Mat Sci Eng C-Mater **33**, 3858-3864 (2013).
37  Rose, J. & Ricciardi, G.: *Statement*, <http://www.formlessfinder.com/statement> (2015).
38  Tsubaki, K.: in *Matter: Material Processes in Architectural Production*  (eds Gail Peter Borden & M. Meredith)  187-203 (Routledge Press, Taylor & Francis Books, 2011).
39  Menges, A.: Material Computation: Higher Integration in Morphogenetic Design Introduction. Architectural Design **82**, 14-21 (2012).
40  Brown, E., Nasto, A., Athanassiadis, A. G. & Jaeger, H. M.: Strain-stiffening in random packings of entangled granular chains. Phys Rev Lett **108**, 108302 (2012).
41  Zou, L. N., Cheng, X., Rivers, M. L., Jaeger, H. M. & Nagel, S. R.: The packing of granular polymer chains. Science **326**, 408-410 (2009).
42  Karayiannis, N. C., Foteinopoulou, K. & Laso, M.: Contact network in nearly jammed disordered packings of hard-sphere chains. Phys. Rev. E **80**, 011307 (2009).
43  Gravish, N., Franklin, S. V., Hu, D. L. & Goldman, D. I.: Entangled Granular Media. Phys. Rev. Lett. **108**, 208001 (2012).
44  Marschall, T. A., Franklin, S. V. & Teitel, S.: Compression- and shear-driven jamming of U-shaped particles in two dimensions. Gran. Matt. **17**, 121-133 (2015).